\documentclass[10pt,tightenlines,eqsecnum,floats,aps,amsmath,
amssymb,nofootinbib,prd,showpacs]{revtex4-1}
\usepackage{amsmath,amsthm,latexsym,amssymb,amsfonts}
\usepackage{enumerate}
\usepackage{graphicx}
\usepackage{calc}
\usepackage{colordvi}
\usepackage[usenames,dvipsnames]{color}
\usepackage{vmargin}
\usepackage{array}

\newcommand{\omicron}{{\cal I}}

\begin{document}

\title{Spin-Foams  for All  Loop Quantum Gravity}

\author{Wojciech Kami\'nski${}^1$, Marcin Kisielowski${}^1$, Jerzy Lewandowski${}^{1,2}$}

\affiliation{${}^1$ Instytut Fizyki Teoretycznej, Uniwersytet Warszawski,
ul. Ho{\.z}a 69, 00-681 Warszawa (Warsaw), Polska (Poland)\\
${}^2$Institute for Gravitation and the Cosmos \&
Physics Department, Penn State, University Park, PA 16802, U.S.A.}

\begin{abstract} \noindent{\bf Abstract\ } The simplicial framework of  Engle-Pereira-Rovelli-Livine  spin-foam models is generalized to match the diffeomorphism invariant framework of  loop quantum gravity. The simplicial spin-foams are generalized to arbitrary linear 2-cell spin-foams. The resulting framework admits all the spin-network states of loop quantum gravity, not only those defined by triangulations (or cubulations). In particular the notion of embedded spin-foam we use allows to consider knotting or linking spin-foam histories. Also the main tools as the vertex structure and  the vertex amplitude are naturally generalized to arbitrary valency case.    
The correspondence between all the SU(2) intertwiners and the SU(2)$\times$SU(2) EPRL intertwiners is 
proved to be 1-1 in the case of the Barbero-Immirzi parameter $|\gamma|\ge 1$, unless the co-domain of the EPRL map is trivial and the domain is non-trivial.
\end{abstract}

\maketitle

\def\be{\begin{equation}}
\def\ee{\end{equation}}
\def\ba{\begin{eqnarray}}
\def\ea{\end{eqnarray}}
\def\lp{{\ell}_{\rm Pl}}
\def\g{\gamma}

\section{Introduction} 
\subsection{Recent spin-foam models of gravitational field}
Spin-foams were introduced as histories of quantum spin-network states of loop quantum gravity  \cite{LQGrevsbooks} (LQG) by Reisenberger and Rovelli \cite{RR}. That idea gave rise to  spin-foam models  \cite{Baezintro,perez,EPRL} (SFM) (see also
Rovelli's book \cite{LQGrevsbooks}). The spin-foam model of  3-dimensional
gravity is derived from a discretization of the BF action (see \cite{Baezintro}
and the references therein). An alternative derivation of the spin-foam model
from $2+1$ LQG was proposed by \cite{NP,perez}. For the 4-dimensional gravity
there were several approaches \cite{RR,Reisenberger}, however, the Barrett-Crane
model \cite{BC} became the obligatory spin-foam model  for almost 10 years.  The
actual relation between that model and LQG  was derived in the seminal works of
Reisenberger \cite{BC_R} and Baez \cite{Baezintro}. The BC model is
mathematically elegant, plays a role in mathematical physics and {\it a priori}
there is no reason for that model to be wrong. Nonetheless, it was shown  not to
have sufficiently many degrees of freedom to ensure the correct classical limit
\cite{BC_error}. That discovery produced an intensive research on suitable
modification. Finally, two teams: $(i)$ Engle, Pereira,
 Rovelli and Livine and $(ii)$  Freidel and Krasnov found  systematic derivations of a spin-foam model of gravitational field using as the starting point a discretization of the Holst action (the Palatini action plus an extra term which does not change the classical
equations of motion is multiplied by an arbitrary real parameter whose inverse is named after Barbero and Immirzi \cite{barbero-immirzi}). The result of their effort is a new, consistent  spin-foam model valid for the Barbero-Immirzi  parameter belonging to the interval $-1\le\gamma\le 1$, a promising candidate for a path integral formulation of 
LQG, and two different models for $|\gamma|> 1$ \cite{flipped,FK,engle-pereira}
(see \cite{engle-pereira} for the comparison). The values of the Barbero-Immirzi parameter predicted by various
black hole models belong to the former interval
\cite{BH}.     

\subsection{The incompatibilities between LQG and some SFMs}
Although the correspondence between LQG and SFMs was settled down recently due to the EPRL model, there are  still several incompatibilities between these frameworks. 

The first difference can be found already at the conceptual level. Loop quantum gravity is a quantum theory of gravitational field with all its local degrees of freedom. This is not a discretized theory. The (kinematical) observables labelled by curves  and 2-surfaces form a complete set for  the continuum theory. The discreteness of LQG is a property of the quantum  representation \cite{LQGdiscr}.  
The SFMs  on the other hand, are derived as quantizations  of discretized classical theories \cite{Reisenberger,EPRL,BD,FC}.     
        
Secondly, LQG is a diffeomorphism invariant theory of fields on manifolds. The
diffeomorphisms play a crucial role in LQG. They are responsible for the diversity
of embedded graphs and spin-networks labelling the quantum states.   
 Theories of discretized spacetime, on the other hand, including the EPRL
model,   seem to describe piecewise flat geometries
defined on piecewise linear  manifolds. In order to match them, either LQG
should be restricted to the piecewise linear 
manifolds  and  piecewise linear spin-networks, or the SFMs should be  suitably generalized. 

The third difference would arise even if one restricted LQG to
the piecewise linear spaces as suggested above. In the LQG canonical framework, whose quantum states are labelled by embedded spin-networks, there is no justified way 
to restrict the graphs to those dual to   triangulations of the underlying
3-manifold. The  EPRL model on the other hand, uses only simplicial complexes 
(or, recently cubulations \cite{thiemann}) and 
the spin-networks defined on their boundaries. Therefore it does not define
a spin-foam history of a generic spin-network state of LQG. In particular,
LQG admits knotted and linked graphs. The simplicial SFMs do not allow such
states as well as they do not allow graphs with vertices more than 4-valent.

In the literature there is a competition between efforts toward piecewise
linearization of LQG \cite{engle,zapata} on the one hand, or generalizations
of the simplicial constructions to the diffeomorphism covariant diversity
\cite{RR,Reisenberger,SF_complex,perez}. We would like to join the latter direction.             

\subsection{The goal of the paper} 
Our aim is to redirect the development of the spin-foam models, and most
importantly the EPRL model,  to that extent, that they can be used to define 
spin-foam histories of an arbitrary spin network state of LQG. 
The notion of embedded spin-foam we use, allows to consider knotting or linking spin-foam histories. Since the knots and links may play a role in LQG, it is an advantage not to miss the chance of keeping those topological degrees of freedom 
in a spin-foam approach. Next, we characterize the structure of a general 
spin-foam vertex. We encode all the information about the vertex structure in the spin-network induced on the boundary of the neighborhood of a given vertex. The converse construction of a vertex from any given embedded spin-network 
was described  by Reisenberger \cite{Reisenberger}. In the set of the 
spin-foams we define the generalized EPRL spin-foams. As an introduction to that
important step, we formulate our definition of the  $n$-valent  Barrett-Crane
intertwiner. That definition is merely equivalent to that of
\cite{BC_R,BC_nvalent,BC_O}, however we spell it out in our notation for the
clarity and the sake of precision. Finally we generalize the
Engle-Pereira-Rovelli-Livine intertwiner to the general spin-foams.    
A technical result concerns the correspondence between all the SU(2) intertwiners and the EPRL intertwiners. We show that the correspondence is 1-1 in the case of the Barbero-Immirzi parameter  $\gamma \ge 1$, unless the co-domain of the EPRL map is trivial and the domain is non-trivial.
   
We were motivated to do this research by two works we want to acknowledge. The
first one is Baez's introduction to SFM \cite{Baezintro}. It is from that
classic paper that we take the definition of linear 2-cell complex with
boundaries and the definition of a spin-foam.  The second paper we appreciate so
much is  
the derivation of the new spin-foam model from the Holst action  by Engle, Livine, Pereira and Rovelli \cite{EPRL}. Those two excellent works brought
spin-foams closer to LQG. On the other hand, the works that should be and will be considered closer in the spirit of the current paper, are the Freidel-Krasnov model \cite{FK} (especially in the range of $\gamma$ in which that model does not overlap with EPRL). Also the pioneering works of Reisenberger \cite{RR,BC_R,Reisenberger}   contain a lot of ideas that still have not been explored enough in the literature.                   
    
\section{The Hilbert space for theories of connections}\label{Sec:connections}
\subsection{The cylindrical functions and the measure}\label{Subsec:cylmu}
Consider a manifold $\Sigma$,  a {compact} Lie group $G$,
its Lie algebra $\mathfrak{g}$, and
the set ${\cal A}(\Sigma)$ of the Lie algebra $\mathfrak{g}$ valued differential one-forms
(connections) on $\Sigma$.

A $G$ valued parallel transport function  on ${\cal A}$ is defined by each
finite curve $p$ in $\Sigma$, namely for every $A\in {\cal A}$,
\be U_{p}(A)\ :=\ {\rm Pexp}\int_p-A\,. \ee
The parallel transport functions are used to define the cylindrical
functions
${\rm Cyl}({\cal A}(\Sigma))$.
A cylindrical function $\Psi:{\cal A}\rightarrow\mathbb{C}$ is defined by
a finite set $P$ of finite, oriented curves $p_1,...,p_n$ in $\Sigma$ and
by a continuous function $\psi : G^n\rightarrow \mathbb{C}$.
 The corresponding cylindrical function is
\be \Psi(A)\ :=\  \psi(U_{p_1}(A),...,U_{p_n}(A)).\ee
The space ${\rm Cyl}({\cal A}(\Sigma))$ of cylindrical functions admits the
sup-norm,
and can be completed to a C$^*$ algebra. Moreover, a
natural diffeomorphism invariant integral is defined
\be {\rm Cyl}({\cal A}(\Sigma))\ \ni \Psi\ \mapsto \ \int d\mu_0(A) \Psi(A). \ee
The idea that led to that integral was the following:  if the differentiability
class of the curves is suitably defined, then for {\it every cylindrical
function} $\Psi\in {\rm Cyl}({\cal A}(\Sigma))$, there is an {\it embedded graph}
$\gamma^{(1)}=\{e_1,...,e_n\}$ and a continuous function $\psi\in{\rm
C}^{(0)}(G^n)$ such that
\be \Psi(A)\ =\ \psi(U_{e_1}(A),...,U_{e_n}(A))\,.\ee
And the continuation of the idea was to {\it define}
the integral as
\be \int d\mu_0(A)\Psi(A)\ :=\ \int d\mu_{\rm H}(g_1)...d\mu_{\rm
H}(g_n)\psi(g_1,...,g_n).\ee
Finally, one had to make sure, that the right-hand side was independent
of the choice of the non-uniquely defined graph $\gamma^{(1)}$. Fortunately,
it was.
The resulting Hilbert space L$^2({\cal A}(\Sigma),\mu_0)$  serves as the Hilbert
space for
background independent quantum theories of (Poisson commuting)
connections, in particular as the kinematical Hilbert space in which
Quantum Geometry of  loop quantum gravity is defined. This is how the
graphs entered the old  loop quantum gravity and gave it the current shape.

The gauge invariant cylindrical functions are easily identified as the
functions given by closed loops. The subspace ${\cal H}_{\Sigma}\subset$L$^2({\cal A}(\Sigma),\mu_0)$ of the gauge invariant functions is defined by
the functions of the form
\be \Psi(A)\ =\ \psi(U_{\alpha_1(A)},...,U_{\alpha_m(A)}),\ee
where $\alpha_1,...,\alpha_m$ are free generators of the {
first} homotopy group of
an embedded graph $\gamma^{(1)}$.

An orthonormal basis of  ${\cal H}_\Sigma$ can be constructed
by endowing the embedded graphs with the spin-network structures {(see
\ref{Subsect:sn})}.

In the Hilbert space L$^2({\cal A}(\Sigma),\mu_0)$ we define a quantum representation
for the classical variables $(A,E)$ where $A$ is the connection  variable introduced above, and $E$ is a $\mathfrak{g}^*$  valued differential form and the Poisson bracket is 
\be \{f,h\}\ =\ \int_\Sigma \frac{\delta}{\delta A^i_a}f\frac{\delta}{\delta E^a_i}h
- \int_\Sigma \frac{\delta}{\delta A^i_a}h\frac{\delta}{\delta E^a_i}f. \ee
In particular, we define the quantum flux operators 
$$\hat{E}(S)\ =\ \frac{1}{i}\int_S \frac{\delta}{\delta A^i_a},$$
across a 2-surface $S\subset \Sigma$.

\subsection{Abstract spin-networks}\label{Subsect:sn}  
\subsubsection{Definition}
\begin{figure}[ht!]
  \centering
    \includegraphics[width=0.36\textwidth]{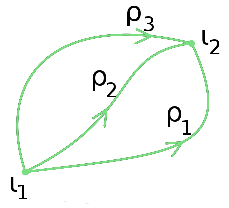}
\caption{A spin-network.}
{\label{sn}}
\end{figure}

Given a  compact group $G$, a $G$-spin-network is a triple
$(\gamma,\rho,\iota)$,  an oriented, piecewise linear 1-complex $\gamma$
(a graph) equipped with two colourings: $\rho$,  and $\iota$ defined
below (see also Fig \ref{sn}).
\begin{itemize}
\item
The colouring $\rho$, maps the set $\gamma^{(1)}$ of the 1-cells (edges)
in $\gamma$ into the set  Irr$(G)$ of the (equivalence classes
of) irreducible  representations of $G$. That is, to every edge $e$ we assign an irreducible representation $\rho_e$ defined in the Hilbert
space ${\cal H}_e$,
\be  e\mapsto \rho_e.\ee

\item The colouring $\iota$ maps each vertex $v\in \gamma^{(0)}$ (where
$\gamma^{(0)}$ denotes the
set of 0-cells of $\gamma$) into the
subspace 
\be {\rm Inv}\left(\bigotimes_{e {\rm\ incoming\ to}\ v}{\rho}^*_e\ \otimes\
\bigotimes_{e'{\rm\ outgoing\ from}\ v}{\rho}_{e'}\right)\  \subset\
\bigotimes_{e {\rm\ incoming\ to}\ v}{\cal H}^*_e\ \otimes\ \bigotimes_{e'{\rm\
outgoing\ from}\ v}{\cal H}_{e'}\label{inv}\ee
each ${\cal H}_e^*$ denotes the dual vector space to ${\cal H}_e$, and
the subspace  consists of all the invariant vectors. The map will
be denoted
\be v\ \mapsto \iota_v. \ee
\end{itemize}
 
\subsubsection{Operations on spin-networks}\label{opsn}
Flipping orientation is the first operation we consider. Given a spin-network
$(\gamma,\rho,\iota)$ let a graph $\gamma'$ be obtained by flipping the
orientation of one of the edges, say $e_1\in \gamma^{(1)}$. The flipped
orientation edge is denoted by $e^{-1}_1$. On $\gamma'$ we can define a
spin-network $(\gamma',\rho',\iota')$ using the data of
$(\gamma,\rho,\iota)$:
\begin{align}
\iota'\ &:=\ \iota\\
\rho'_e\ &:=\ \begin{cases}(\rho_{e_1})^*,\ \  e=e_1^{-1}\\
                            \rho_e, \ \ {\rm
otherwise}.\end{cases}\label{flip}
\end{align}

The complex conjugate spin-network $(\gamma,\overline{\rho},\overline{\iota})$  to a given spin-network $(\gamma,\rho,\iota)$ is defined by
using the conjugation
map of any vector space
$$V\ni v\mapsto \bar{v}\in \bar{V}.$$
The conjugate vector space  $\bar{V}$ is defined as the same set $V$ with
a new multiplication $\bar{\cdot}$ defined to be 
$$  a\,\bar{\cdot}\,v\ :=\ \bar{a}v, $$
the same adding operation $\bar{+}=+$, and the conjugation map being the identity.
Clearly,
\begin{align}
\bar{\iota}_v\ := \overline{\iota_v},\ \ \ \ \ \bar{\rho}_e\ :=\
\overline{\rho_e}.
\end{align}

The Hilbert conjugate spin-network $(\gamma^\dagger,
\rho^\dagger, \iota^\dagger)$ to a  spin-network
$(\gamma,\rho,\iota)$ is a spin-network defined on the graph $\gamma^\dagger$  obtained by flipping the orientation of each
of the edges of $\gamma$,
 and
\begin{align}\rho^\dagger_{e^{-1}}\ &:=\ \rho_e, \\
             \iota^\dagger_v\ &:=\ (\iota_v)^\dagger \end{align}
where given the Hilbert space ${\cal H}$, we denote by
$${\cal H}\ni v\ \mapsto\ v^\dagger\in {\cal H}^*$$
the antilinear map  defined by the Hilbert product (that is ``
$|v\rangle^\dagger=\langle v|$  ''). It is not hard to check, though,  that
each spin-network $(\gamma^\dagger,\rho^\dagger,\iota^\dagger)$ can be obtained from
the complex conjugate spin-network
$({\gamma},\bar{\rho},\bar{\iota})$ by the operations of flipping
orientation of each of the edges of $\gamma$. 

Splitting an edge of  a spin-network $(\gamma,\rho,\iota)$ consists of considering
the graph $\gamma'$ obtained from $\gamma$ by splitting one of its edges, say $e$ into
$$ e\ =\ e'_2\circ e'_1,$$
where we use a convention, that one first runs through $e'_1$ and then
through $e'_2$. The edges $e'_1,e'_2$ are oriented in the
agreement with $e$ and we insert the identity
into a new vertex $v_{12}$ connecting the edges $e'_1$ and $e'_2$. In detail, 
\begin{align}
\rho'(e')\ &:=\ \begin{cases}\rho(e),\ \ {\rm if}\ e' = e'_1,e'_2\\
                            \rho(e')\ \ {\rm if}\ e'\in \gamma^{(1)}  
                            \end{cases}\\
\iota'(v')\ &:=\ \begin{cases} {\rm id},\ \ {\rm if}\ v'=v_{12}\\
                                  \iota(v')\ \ {\rm if}\ v'\in\gamma^{(0)} \end{cases}                            
 \end{align}

\subsubsection{Spin-network functions}
With a spin-network $s=(\gamma,\rho,\iota)$ we associate a
function\footnote{Given two sets $X$ and $Y$, by $Y^X$ we denote the set
of maps
$X\rightarrow Y$. If $X$ has $n$ elements then $Y^X\sim Y^n$. This
notation lets us avoid choosing an ordering in $X$.}
\begin{align}
\psi_{s}\ :\ G^{\gamma^{(1)}}\ &\rightarrow\ \mathbb{C}.
\end{align}
Denote elements of $G^{\gamma^{(1)}}$ by
$$ g:\gamma^{(1)}\ \rightarrow\  G,\ \ \ e\mapsto g_e.$$
For every $g\in  G^{\gamma^{(1)}}$ there is a unique contraction
\be \psi_s(g)\ :=\  \left(\bigotimes_{e\in \gamma^{(1)}}
\rho_e(g_e)\right) \lrcorner
\left(\bigotimes_{v\in\gamma^{(0)}}\iota_v\right).\label{psis}\ee
In the abstract index notation, the contraction is defined as follows. For
every edge $e\in\gamma^{(1)}$ we have $\rho^A_B(g_e)$ in (\ref{psis}).  At
the beginning point $v$ of $e$, there  is an invariant
$\iota_{v}{}_{...A...}{}^{...}$ in (\ref{psis})
(the dots stand for the remaining indices) and at the end point $v'$ of
$e$, there is an invariant $\iota_{v'}{}_{...}{}^{...B...}$. The
corresponding part of (\ref{psis})
reads
$$...\,\iota_{v}{}_{...A...}{}^{...}\,\rho^A_B(g_e)\,\iota_{v'}{}_{...}{}^{...B...}\,...$$

Given a graph $\gamma$, the spin-network functions form naturally a
basis of the Hilbert space ${\cal H}_\gamma\subset
L^2(G^{\gamma^{(1)}},\mu_{\rm H})$, where $\mu_{\rm H}$ is the Haar
measure. The subspace ${\cal H}_\gamma$ coincides with the subspace
of gauge invariant elements of L$^2(G^{\gamma^{(1)}},\mu_{\rm H})$, where
the gauge transformations are defined as follows:
given a vertex $v\in\gamma^{(0)}$   the gauge transformation defined by
$h\in G$ in $G^{\gamma^{(1)}}$ is
\be (h,v)g_e\ =\ \begin{cases} g_eh, \ {\rm if\ }e\ {\rm begins\ at\ }v\
{\rm and\ ends\ elsewhere}\\
                           h^{-1}g_e, \ {\rm if\ }e\ {\rm ends\ at\ }v\
{\rm and\ begins\ elsewhere}\\
                           h^{-1}g_eh,\ {\rm if\ }e\ {\rm begins\ and\
ends\ at\ }v\\
                            g_e,\ {\rm otherwise.}\end{cases}\ee
The general gauge transformation is defined by a sequence of elements of
$G$ labelled by vertices.

Next, if a graph $\gamma'$ is obtained by flipping the orientation 
in one of the edges, say $e_0$, of $\gamma$, then we define a map
\begin{align} \label{map1}
 G^{\gamma^{(1)}}\ &\rightarrow G^{\gamma'^{(1)}},\\ 
g_e\ &\mapsto\ \begin{cases} (g_{e_0^{-1}})^{-1},\ {\rm if}\ e=e_0\\
                              g_e,\ {\rm otherwise}\end{cases} 
\end{align}
  
In the case when $\gamma'$ is obtained by splitting one of the edges, say $e_0$, of
$\gamma$ into $e=e'_2\circ e'_1$, the corresponding map is 
\begin{align} \label{map2} f:G^{\gamma'^{(1)}}\ &\rightarrow G^{\gamma^{(1)}},\\ 
f(g')_e\ &=\ \begin{cases} g_{e'_2}g_{e'_1},\ {\rm if}\ e=e_0\\
                              g'_e,\ {\rm otherwise}\end{cases}. 
\end{align}
\subsubsection{Equivalence relation} 
The spin-network functions define a natural equivalence relation in the space
of spin-networks, namely two spin-networks are equivalent if and only if
the corresponding spin-network functions  are equal modulo the operations
of splitting edges and/or reorienting edges accompanied by the maps
(\ref{map1}) and, respectively, (\ref{map2}). Going back to the operations on
the spin-networks of Sec. \ref{opsn}, we can see
that splitting an edge of a spin-network and flipping the orientation map a spin-network into an equivalent one. Not surprisingly, the complex conjugation of a spin-network      
has the following interpretation in terms of the spin-network functions
\be \psi_{\bar{s}}\ =\ \overline{\psi_s}. \ee

The conclusion is, that modulo the equivalence relation we can 
change the orientation or split edges, not changing a given spin-network.
This definition is also useful to define an equivalence of spin-foams. 

\subsubsection{Evaluation of a spin-network}
Evaluation of a spin-network $s=(\gamma,\rho,\iota)$ is the number
\be s\ \mapsto\ \psi_s(I),  \ee
where $I\in G^\gamma$ is the identity element.

\subsection{The spin-network cylindrical functions of connections}\label{Subsec:sncyl}
Finally, we are in a position  to explain the application of the
spin-networks
in the Hilbert space of the cylindrical functions introduced in  the
previous section.  Consider  the 1-complexes (graphs) {\it embedded} in
$\Sigma$, and embedded spin-networks defined on them.
Given an embedded spin-network $s=(\gamma,\rho,\iota)$, we use the
corresponding
spin-network function $\psi_s$, to define a spin-network cylindrical
function $\Psi_s:{\cal A}\rightarrow \mathbb{C}$. First,  a connection
$A\in{\cal A}$ defines an element $U_\gamma(A)\in G^{\gamma^{(1)}}$,
\be U_\gamma(A) :\ e\ \mapsto\ U_e(A)\in G.\ee
Next, we use the spin-network function $\psi_s$,
\be \Psi_s(A)\ :=\ \psi_s(U_\gamma(A)).\ee

If we put a restriction on embeddings to be piecewise analytic then the
spin-network cylindrical functions span a dense subset of the Hilbert
space
${\cal H}_\Sigma$ and it is easy to construct from them an orthonormal basis.
This assumption allow us to overcome some degenerated behavior of the smooth
category.

The cylindrical functions lead to a natural equivalence relation between
embedded spin-networks: two embedded spin-network states are equivalent 
if and only if the corresponding spin-network states are equal.   

\section{Spin-foams of spin-networks and of spin-network functions}\label{Sec:sf}
\subsection{Classical motivation}\label{Subsec:sfmotiv}  
The motivation of the spin-foam approach to LQG is to develop an analog of
the Feynman path integral. The idea of Rovelli and Reisenberger \cite{RR}, is 
that the paths, should be suitably defined histories of the spin-network states.
We address this issue in this section.   
\begin{figure}[ht!]
  \centering
    \includegraphics[width=0.5\textwidth]{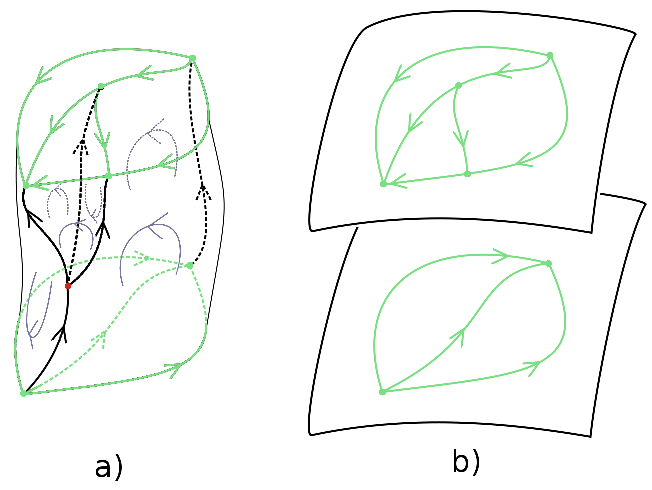}
\caption{a) A history of a spin-network. b) The initial, and, respectively,
final spin-network.}
\end{figure}

\subsection{Spin-foams}\label{Subsec:sf}    
\subsubsection{Foams}\label{Subsubsec:f} 
By a foam we mean throughout this work an oriented linear 2-cell complex
with (possibly empty) boundary. For the precise definition of the linear cell complexes we refer the reader to \cite{Baezintro,pl}. Briefly, each foam $\kappa$ consists of 2-cells (faces), 1-cells (edges), and 0-cells
(vertices).
\begin{figure}[ht!]
  \centering
    \includegraphics[width=0.4\textwidth]{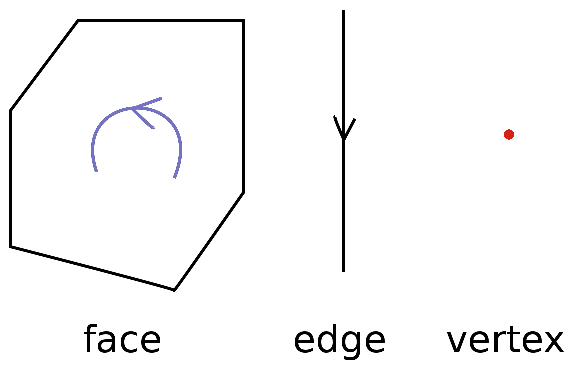}
\caption{{Cells of the complex.}}
{\label{fev}}
\end{figure}

 The faces are  polygons, their sites are edges,  the ends of
the  edges are vertices. Faces and edges are oriented, and the orientation
of an edge
is independent of the orientation of the face it is contained in. Each edge
is contained in several (at least one) faces, each vertex is contained in
several
(at least one) edges.

The boundary $\partial \kappa$ is a 1-cell  subcomplex (graph)
of $\kappa$
{ such that there exists a one-to-one affine map $c:\partial
\kappa \times [0,1] \to \kappa$ which maps each cell of $\partial
\kappa \times [0,1]$ onto the unique cell of $\kappa$ and the set
$\partial \kappa \times [0,1[$ onto an open subset of $\kappa$ (see appendix of \cite{Baezintro} for details).
An  edge of $\kappa$ is an edge of the boundary if and
only
if it is contained in $\partial \kappa$. Otherwise, it is  an internal edge.} A
vertex of $\kappa$ is a vertex of $\partial \kappa$ if and only if it is
contained in exactly one internal edge of $\kappa$ (this is an important
technical subtlety of the definition of a boundary). Otherwise, it is an
internal vertex of $\kappa$.

\subsubsection{Colouring}\label{Subsubsec:col}
Given a foam  $\kappa$, a spin-foam is defined by introducing  two
colourings:
\begin{itemize}
\item  $\rho$ colours  the faces of $\kappa$   with irreducible
representations of the group $G$ (the set of the faces of a foam $\kappa$
is denoted by $\kappa^{(2)}$),
\ba \rho:\ \kappa^{(2)}\ &\rightarrow\ {\rm Irr}(G),\\
                      f\ &\mapsto\ \rho_f
\ea
We consider representation $\rho_f$ as acting on a given Hilbert space ${\cal
H}_f$.


\item $\iota$ colours the internal edges of $\kappa$ (their set is denoted
by $\kappa^{(1)}_{\rm int}$) with invariants of suitable tensor product of
the representations given  by the colouring $\rho$. Let $e\in
\kappa^{(1)}_{\rm int}$. To define the space of invariants Inv$_{e}$  we split the set of faces
containing $e$, into faces whose orientation coincides with that of $e$,
and, respectively, with the opposite orientation,
\be {\rm Inv}_e\ \subset\ \bigotimes_{f\ {\rm same\ orientation}}{\cal
H}_f\,\otimes\,\bigotimes_{f'\ {\rm opposite\ orientation}}{\cal
H}^*_{f'},\ee
where the subset consist of the invariants of the representation
\be \bigotimes_{f\ {\rm same\ orientation}}{\cal
\rho}_f\,\otimes\,\bigotimes_{f'\ {\rm opposite\ orientation}}{\cal
\rho}^*_{f'}.\ee
The colouring $\iota$ is a map
\be e\ \mapsto\ \iota_e\ \in\ {\rm Inv}_e .\ee
\end{itemize}
\begin{figure}[ht!]
  \centering
    \includegraphics[width=0.6\textwidth]{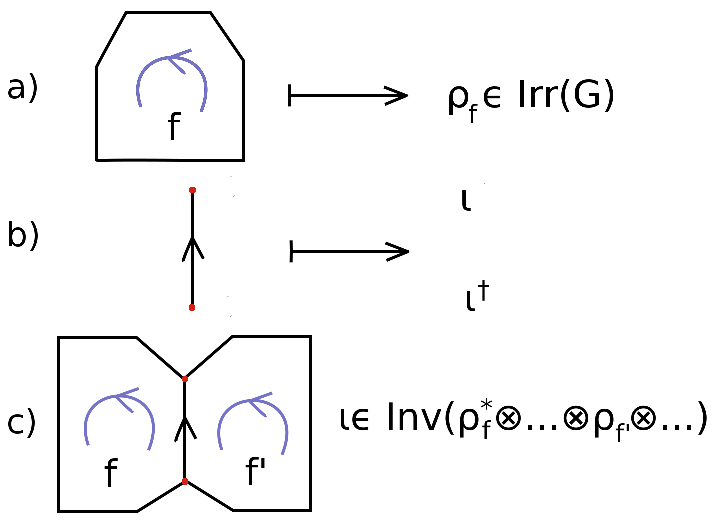}
\caption{{a) }Faces are coloured by irreducible representations of $G$.
{b), c) }Edges are
coloured with invariants. It is convenient to assign $\iota_e^\dagger$ to the
beginning of $e$ and $\iota_e$ to the end of $e$.}
{\label{colour}}
\end{figure}
Note that the spin-foam structure has been defined in the interior of a
foam.

In fact, it is often convenient to think of $\iota_e^\dagger$ as assigned to the
edge $e$ at the beginning point whereas at the end point we assign
$\iota_e$.

\subsubsection{Induced boundary spin-network}\label{Subsubsec:indsn}

Given a spin-foam $(\kappa,\rho,\iota)$, the colourings $\rho$ and $\iota$
induce on the boundary $\partial \kappa$ a spin-network structure
$(\partial \kappa, \partial \rho,
\partial \iota)$. For every edge $e$ of $\partial\kappa$ let $f_e$ denote
the {unique} face of $\kappa$ which contains $e$, and
\begin{equation}
\partial \rho_e\ :=\ \begin{cases}
\rho_{f_e}, \ \ {\rm if\ the\ orientations\ of}\ f_e\ {\rm and}\ e\ {\rm
coincide}\\
\rho_{f_e}^*,\ \ {\rm if\ they\ are\ opposite}.
 \end{cases}
\end{equation}
For every vertex $v$ of $\partial \kappa$, let $e_v$ be the {
unique} internal edge
of $\kappa$ which contains $v$, and
\begin{equation}
\partial \iota_v\ :=\ \begin{cases}
\iota_{e_v}^\dagger, \ \ {\rm if}\ v \ {\rm is\ the\ beginning\ of}\ e_v\\
\iota_{e_v}\ \ {\rm if}\ v \ {\rm is\ the\ end\ of}\ e_v.
\end{cases}\end{equation}
\begin{figure}[ht!]
  \centering
    \includegraphics[width=0.9\textwidth]{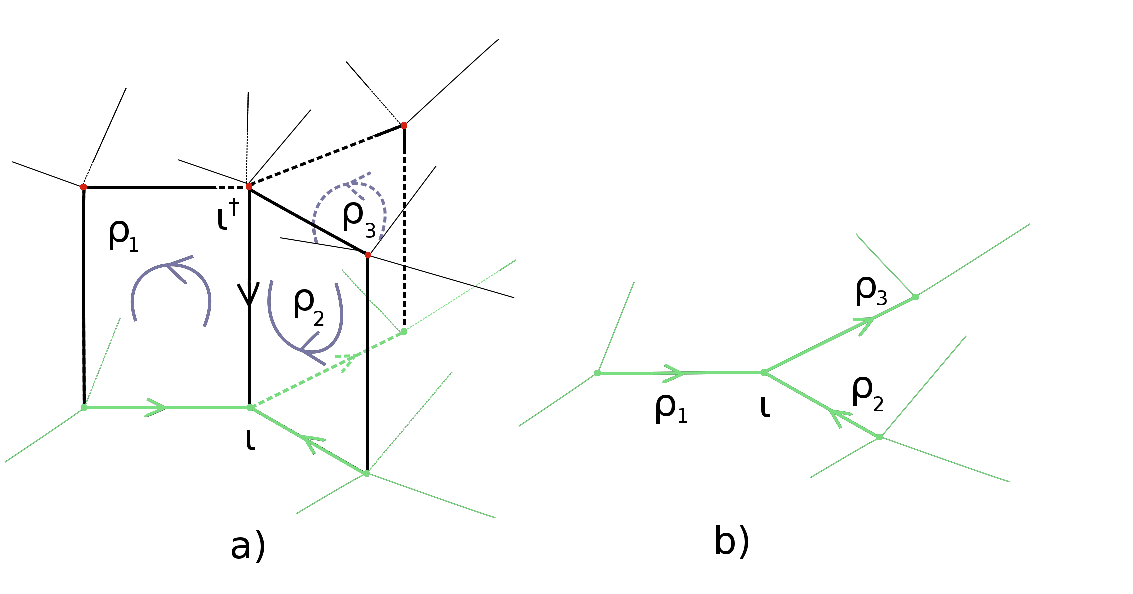}
\caption{a) Spin-foam with boundary (in the bottom). b) Induced spin-network on the boundary.}
\end{figure}
\subsubsection{Operations on spin-foams}\label{Subsubsec:opsf} 
Given a spin-foam
$(\kappa,\rho,\iota)$ consider an oriented complex  $\kappa'$, obtained by flipping the orientation of one of the faces, say $f\in \kappa^{(2)}$. Denote the flipped
orientation face by $f^{-1}$. On $\kappa'$ we can define a
spin-foam $(\kappa',\rho',\iota)$ modifying $\rho$,
\begin{equation} 
\rho'_{f'}\ :=\ \begin{cases}(\rho_{f})^*,\ \  f'=f^{-1}\\
\rho_{f'}, \ \ {\rm
otherwise}.\end{cases}\label{flips1}
\end{equation}
 
Another operation involving a change of the orientation of a given spin-foam
$(\kappa,\rho,\iota)$ is flipping the orientation of an edge $e$. Denote the 
resulting foam by $\kappa'$. If $e$ is a boundary edge, we just leave the labellings $\rho$ and $\iota$ unchanged and consider the spin-foam $(\kappa',\rho,\iota)$. If 
$e$ is an internal edge, we modify $\iota$ as follows
 \begin{equation} 
\iota'_{e'}\ :=\ \begin{cases}(\iota_{e})^\dagger,\ \  e'=e^{-1}\\
                            \iota_{e'}, \ \ {\rm
otherwise}.\end{cases}\label{flips}
\end{equation}

In both cases presented above, (\ref{flips1}) or (\ref{flips}), the resulting 
spin-foam will be considered equivalent to the starting $(\kappa,\rho,\iota)$. 
This equivalence allows to reorient spin-foams in the same class of equivalence.
 
Yet another operation is splitting a spin-foam $(\kappa,\rho,\iota)$,  into two spin-foams $(\kappa_1,\rho_1,\iota_1)$ and $(\kappa_2,\rho_2,\iota_2)$ such that:
\begin{itemize}
\item $\kappa_1,\kappa_2$ are subcomplexes of $\kappa$
\item the common part of $\kappa_1$ and $\kappa_2$ is a 1-subcomplex
of the boundary of $\kappa_1$, as well as of the boundary of $\kappa_2$
\end{itemize}
The inverse operation operation to the splitting is called gluing of spin-foams.

\subsubsection{Spin-foam equivalence}\label{Subsubsec:sfequiv}
The equivalence relation introduced in the space of the spin-networks can be used
to introduce equivalence relation between spin-foams. Spin-foams $(\kappa,\rho,\iota)$
and $(\kappa',\rho',\iota')$ are equivalent if the following two conditions
are satisfied:
{
\begin{itemize}
\item there exists a bijective affine map (possibly not preserving orientation
of the cells) from $\kappa$ onto $\kappa'$
\item  for every splitting of $(\kappa,\rho,\iota)$ and the related
splitting of $(\kappa',\rho',\iota')$, the corresponding spin-networks induced
on the boundaries are equivalent.
\end{itemize}
}

\subsection{Spin-foam trace}\label{Subsec:sftr}
\subsubsection{The vertex trace}\label{Subsubsec:vtr} To every internal vertex  of a spin-foam $(\kappa,\rho,\iota)$ we can naturally assign a number  by contracting the
invariants 
$\iota_e\,/\,\iota_{e'}^\dagger$ which colour the incoming/outgoing
edges. To
begin with, given an internal  vertex $v$  consider the tensor product
\begin{equation}\label{iotas} \bigotimes_{{\rm outgoing}\ e} \iota_{e}^\dagger\
\otimes\  \bigotimes_{{\rm incoming}\ e'} \iota_{e'}
 \end{equation}
For every face $f$ intersecting $v$, there are exactly two edges at $v$,
say $e_1$ and $e_2$, contained in $f$. Take the corresponding invariants
present in  (\ref{iotas}). One of them has exactly one index corresponding
to the representation $\rho_f$. Then, the other one has exactly one index in the  representation $\rho_f^*$. Contract those indices. And repeat the procedure for every face intersecting $v$.
The result can be symbolically denoted by,
   \be {\rm Tr}\left( \bigotimes_{{\rm outgoing}\ e} \iota_{e}^\dagger\ \otimes\
\bigotimes_{{\rm incoming}\ e'}
\iota_{e'}\right)\label{trtensoriotas}\ =: {\rm Tr}_v(\kappa,\rho,\iota)\ee
and we call it the spin-foam vertex trace at a vertex $v$.
\begin{figure}[ht!]
  \centering
    \includegraphics[width=0.8\textwidth]{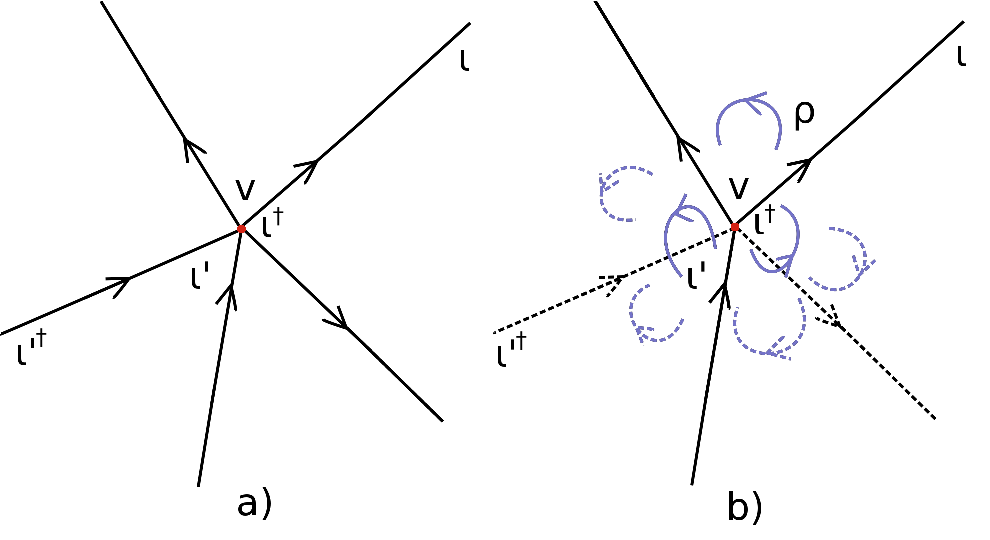}
\caption{a) Edges meet at vertices. Every edge contributes an invariant or
hermitian conjgate invariant. b) Every face meeting $v$ contains
exactly two of the edges.}
{\label{vrtx1}}. 
\end{figure}
\begin{figure}[ht!]
  \centering
    \includegraphics[width=0.8\textwidth]{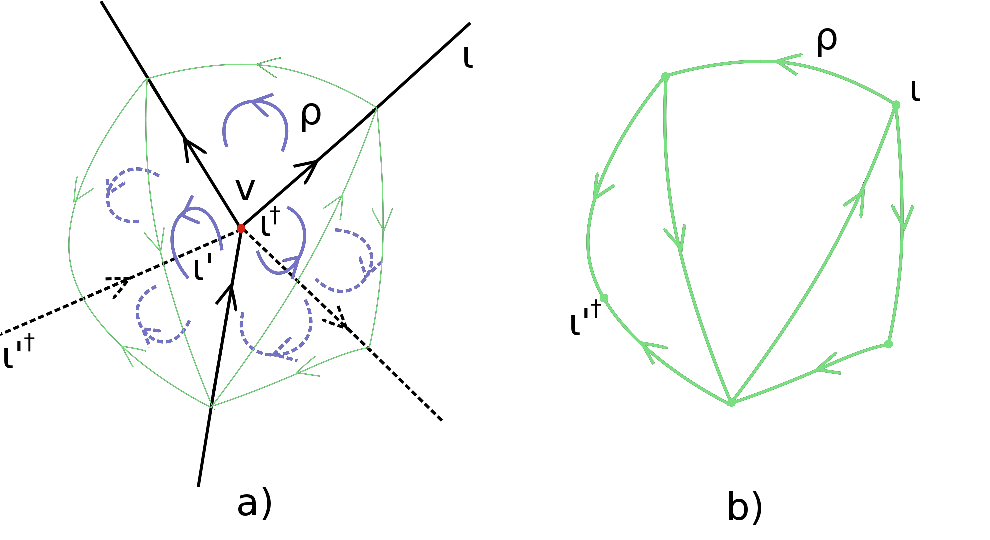}
\caption{a) A foam neighbourhood of the vertex $v$ bounded by  new, thin edges.
b) The spin-network induced on the boundary of the neighbourhood. We call it:
the vertex
spin-network.}
{\label{vrtx}}
\end{figure}

\subsubsection{The vertex spin-network} The spin-foam vertex trace has a 
clear  interpretation in terms of the spin-networks. Consider a spin-foam 
$(\kappa,\rho,\iota)$.
Given an internal vertex $v$ and the intersecting faces, on each of the faces
we consider a suitable neighbourhood of $v$. The neighbourhood is 
bounded by the segments of the sites of the face meeting at $v$, and a new
extra edge connecting the segments (marked  by a thinner curve on Fig. \ref{vrtx} a)). The union of the resulting neighbourhoods 
of $v$ in each intersecting face is a foam neighbourhood of the vertex $v$. The foam neighbourhood of $v$  is a spin-foam with boundary itself. The boundary is formed by the thinner edges on Fig. \ref{vrtx} a). Denote the spin-network induced  on the boundary (Fig. \ref{vrtx} b)) by $s$, and the corresponding spin-network
function by $\psi_s$. The vertex trace  (\ref{trtensoriotas}) of the spin-foam equals the evaluation of the conjugate spin-network
\be \overline{\psi_s(I)}\ =\ {\rm Tr}_v(\kappa,\rho,\iota). \ee
The conjugate spin-network itself is the spin-network induced on the boundary of
the spin-foam obtained from $(\kappa,\rho,\iota)$ by removing the neighbourhood of the vertex $v$.
  
\subsubsection{The spin-foam trace} The spin-foam trace of a spin-foam $(\kappa,\rho,\iota)$  will be defined by the product
of the internal vertex traces. For the consistency with respect to  the gluing,  we introduce  a  factor for every vertex $v$ of the boundary of the spin-foam,
\be  \sqrt{{\rm Tr}(\partial\iota_v^\dagger\partial \iota_v)}\ee    
where $(\partial \kappa, \partial\rho,\partial\iota)$ is the spin-network induced
on the boundary {and the Tr is the contraction of the interrelated
indices}.
The spin-foam trace is defined to be
\be {\rm Tr}(\kappa,\rho,\iota)\ :=\ \prod_{{\rm internal\ vertices}\ v} {\rm Tr}_v(\kappa,\rho,\iota)\prod_{{\rm boundary\ vertices}\ v} \sqrt{{\rm Tr}(\partial\iota_v^\dagger\partial \iota_v)}\label{sftr}\ee 

If a spin-foam $(\kappa,\rho,\iota)$ is the result of the  gluing of spin-foams
 $(\kappa_1,\rho_1,\iota_1)$, and $(\kappa_2,\rho_2,\iota_2)$ then
 \be {\rm Tr}(\kappa,\rho,\iota)\ =\ {\rm Tr}(\kappa_1,\rho_1,\iota_2){\rm Tr}(\kappa_2,\rho_2,\iota_2). \ee 
Indeed, if a boundary vertex $v_1$ in $(\kappa_1,\rho_1,\iota_1)$ and a boundary vertex $v_2$ in $(\kappa_2,\rho_2,\iota_2)$ after the gluing become a single internal vertex $v$ of    $(\kappa,\rho,\iota)$, then 
\be {\rm Tr}_v (\kappa,\rho,\iota)\ =\ {\rm
Tr}(\partial\iota_{v_1}^\dagger\partial \iota_{v_1})\ =\ {\rm
Tr}(\partial\iota_{v_2}^\dagger\partial \iota_{v_2}).
\ee

\subsection{Spin-foam of a spin-network function}\label{Subsec:sfsnf} Thus far in this section
we have been considering abstract spin-networks and  abstract spin-foams.
As we remember from the previous section, every spin-network state $\Psi_s$ is an element of the Hilbert space ${\cal H}_{\Sigma}$ of  gauge invariant elements of the Hilbert space L$^2({\cal A}(\Sigma),d\mu_0)$ of the cylindrical functions of $G$-connections on $\Sigma$. The spin-network state is defined by a spin-network $s$ embedded in a  manifold  $\Sigma$. 

There are two settings for considering spin-foams of spin-network states: 
\begin{enumerate}
\item The causal approach \cite{Markopoulou,Baezintro,perez} within which a spin-foam  $(\kappa,\rho,\iota)$ is a
history of an initial spin-network state $\Psi_{s_{\rm in}}\in {\cal H}_{\Sigma_{\rm in}}$ defined by a spin-network $s_{\rm in}$ embedded in a manifold $\Sigma_{\rm in}$. The history  ends at a final spin-network state $\Psi_{s_{\rm out}}\in {\cal H}_{\Sigma_{\rm out}}$ defined by a spin-network $s_{\rm out}$ embedded in $\Sigma_{\rm out}$. The histories are defined by foams $\kappa$ embedded in a given 
manifold $M$ whose boundary is the disjoint sum  
$$\partial M\ =\ \Sigma_{\rm in}\cup\Sigma_{\rm out},$$
and 
$$ \partial \kappa\ \subset \Sigma_{\rm in}\cup\Sigma_{\rm out}.$$
For each  of the embedded spin-foams $(\kappa,\rho,\iota)$, { the
spin-network $s_{\rm in}$ is the hermitian adjoint
($=$ the complex conjugate) of the spin-network induced on  $\partial
\kappa\subset\Sigma_{\rm in}$ whereas $s_{\rm out}$ is the spin-network
induced on $\partial
\kappa\subset\Sigma_{\rm out}$}(see explanation
below). 

\item  The surface approach  \cite{LQGrevsbooks} (see Rovelli's book) and \cite{graviton,RovQM}, within which one considers  a manifold    $M$
and spin-network states $\Psi_s\in {\cal H}_{\Sigma}$ defined by the spin-networks embedded in 
$$\partial M\ =\ \Sigma.$$
This approach is particularly natural in the Euclidean theory. 
\end{enumerate} 

The emergence of the complex conjugation in the first item above, can be understood on the simplest example of  the trivial spin-foam evolution (see Fig. [8]). Consider 
 the product manifold 
 $$M\ =\ \Sigma\times [0,1]\ , $$ 
and  an embedded spin-network 
 $$s_{\rm in}\ =\ (\gamma,\rho,\iota)\ $$
{The spin-network $s$ induced on the boundary}
$\Sigma\times \{0\}$ { is obtained from $s_{\rm in}$ by flipping the
orientations of the edges and taking the hermitian adjoint of its intertwiners,
hence, $$s\ =\ (\gamma^\dagger,\rho^\dagger,\iota^\dagger)\ $$ }.
Construct the ``static'' spin-foam history.
The spin-foam faces are 
\be f=e\times [0,1]\ ,\ee where $e$ runs through the set of the edges of $\gamma$, and the spin-foam edges are 
\be v\times  [0,1]\ ,\ee where $v$ runs through 
the set of the vertices of $\gamma$. Orient each  face $e\times [0,1]$ consistently with the edge $e$ and each edge $v\times  [0,1]$ according to the
interval. Finally, the spin-network $s'$
induced on the boundary $\Sigma\times\{1\}$ 
is obtained from $s$ by flipping the orientations of the edges and taking the hermitian adjoint of its intertwiners, hence,  
{\be s'\ =\ (\gamma,\rho,\iota)\ ,\ee}
according to the definitions from Sec. \ref{opsn}, hence the corresponding state
is
\be \Psi_{s'}\ =\ \overline{\Psi_{s_{\rm in}}},\ee
provided we identify $\Sigma\times\{0\}$ with  $\Sigma\times\{1\}$. According  to the definition of the final spin-network state 
{\be s_{\rm out}\ =\ s'\ =\ (\gamma,\rho,\iota)\ =\ s^\dagger=
s_{\rm in}, \ee}
in this static spin-foam case, as one would expect. 
\begin{figure}[ht!]
  \centering
    \includegraphics[width=0.9\textwidth]{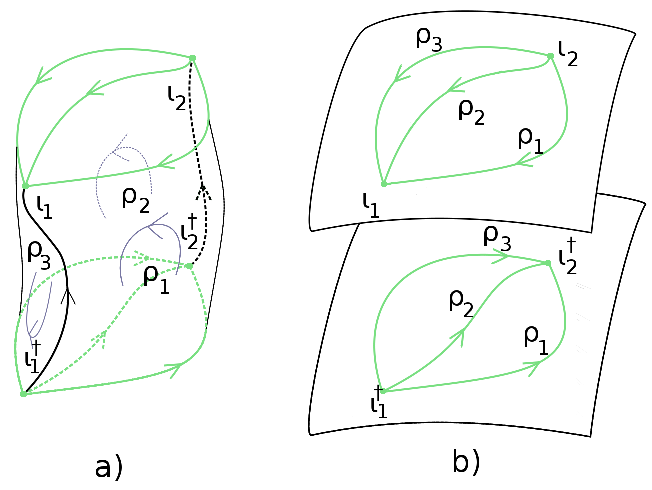}
\caption{a) The static history of a spin-network state.  b) The spin-network induced
on the {lower } part of the boundary is the
hermitian adjoint of the the initial (and 
at the same time final)  spin-network.} 
\end{figure}
There is a natural equivalence relation in the space of embedded spin-foams in a
given $\Sigma\times \mathbb{R}$. Given an embedded spin-foam
$(\kappa,\rho,\iota)$, a generic cross section 
of $\Sigma\times\mathbb{R}$ cuts the spin-foam into two spin-foams with boundaries,
and the spin-foam induces an embedded spin-network $s$ on the cross section (say, that we use the upper part of the spin-foam to induce $s$) and the corresponding spin-network state $\Psi_s$. Another embedded spin-foam is equivalent to $(\kappa,\rho,\iota)$ if it induces the same spin-network state on every generic cross section.  
            
\section{The Spin-foam models of Euclidean 4-spacetime}\label{Sec:sfmodels} 
\subsection{Motivation}\label{Subsubsec:sf4dmot} The starting point is the gravity theory in terms of the Palatini action 
\be S_{\rm P}(e,A)\ =\ \frac{1}{2\kappa} \int_{\Sigma\times \mathbb{R}}\frac{1}{2}\epsilon_{IJKL} e^I\wedge e^J \wedge F^{KL},   \ee
where $(e^1,e^2,e^3,e^4)$ is an orthonormal tetrad of 1-forms,
$(A^1_1,A^2_1,...,A^4_4)$ are connection 1-forms which set an antisymmetric 1-form matrix 
\be A^K_J \eta_{KI} + A^K_I\eta_{KJ}\ =\ 0,\ee  
of the curvature
\be F^I_J\ =\ dA^I_J\ +\ A^I_K\wedge A^K_J\ee
and the indices are raised and lowered with the matrix $\eta_{IJ}$ which in the Euclidean case is $\delta_{IJ}$. A classically irrelevant modification 
whose result is called the Holst action \cite{Holst} and reads
\be S_{\rm H}(e,A)\ =\ \frac{1}{2\kappa} \int_{\Sigma\times \mathbb{R}}\frac{1}{2}\epsilon_{IJKL} e^I\wedge e^J \wedge F^{KL}\ +\ \frac{1}{\gamma}e^I\wedge e^J \wedge F_{IJ}, \ee 
in fact leads to nonequivalent quantum theory depending on the value of the Barbero-Immirzi parameter $\gamma$.  The theory can be interpreted as given by the action
\be S(E,A)\ =\  \int_{\Sigma\times \mathbb{R}} E_{IJ}\wedge F^{IJ}\label{BF} \ee      
with additional simplicity constraint     
\be  E_{IJ}\ =\ \frac{1}{2}\epsilon_{IJKL}e^I\wedge e^J    \ee 
in the Palatini case and
\be  E_{IJ}\ =\ \frac{1}{2}\epsilon_{IJKL}e^K\wedge e^L\ +\ 
\frac{1}{2\gamma}e_I\wedge e_J,   \label{simpl} \ee 
in the Holst action case. 

\subsection{The scheme for embedded spin-foam models}\label{Subsec:sch}
The recent spin-foam models of Barrett-Crane \cite{BC},
Engle-Pereira-Rovelli-Livine \cite{EPRL}
and Freidel-Krasnov \cite{FK} fall into a common scheme.  We combine it in below with the assumption that the spin-foams are histories of the embedded spin-network functions defined within the framework of the diffeomorphism invariant quantization
of the theory of connections (see Section \ref{Sec:connections}).  The resulting scheme reads as follows:  

\begin{itemize}

\item First, ignoring the simplicity constraints, one quantizes the theory (\ref{BF})
using the Hilbert space L$^2({\cal A}^+(\Sigma),{\rm d}\mu_0)\times$L$^2({\cal A}^-(\Sigma),{\rm d}\mu_0)$ of the cylindrical functions 
of the SU(2)$\times$SU(2) connections on a  3-surface $\Sigma$ and the subspace
${\cal H}^+_\Sigma\times{\cal H}^-_\Sigma$ of the gauge invariant elements.

\item Next, the simplicity constraints are suitably quantized to become linear quantum constraints imposed on the elements of ${\cal H}_\Sigma$. 
The constraint is locally defined at every vertex, hence it reduces to a constraint defined in each space ${\rm Inv}(\rho_1\otimes...\rho_k\otimes\rho^*_{k+1}\otimes...\otimes\rho^*_{N})$
of the intertwiners. It defines a subspace ${\rm Inv}_{\rm
Simp}(\rho_1\otimes...\rho_k\otimes\rho^*_{k+1}\otimes...\otimes\rho^*_{N})$ of
{\it simple intertwiners}.   
That subspace may be  nonempty only for some subset of representations. The spin-network functions defined by the simple intertwiners
span the space of solutions to the simplicity constraint in  ${\cal H}_\Sigma$.

\item  We consider the spin-foams  of spin-network functions  introduced in Section \ref{Subsec:sfsnf} with the intertwiners restricted to the spaces ${\rm Inv}_{\rm Simp}(\rho_1\otimes...\rho_k\otimes\rho^*_{k+1}\otimes...\otimes\rho^*_{N})$ 
of simple intertwiners. That is, for every spin-foam $(\kappa,\rho,\iota)$ we consider, the map $\iota$ takes values in corresponding
subspaces ${\rm Inv}_{\rm Simp}(\rho_1\otimes...\rho_k\otimes\rho^*_{k+1}\otimes...\otimes\rho^*_{N})$.
  
\item The amplitude of each spin-foam $(\kappa,\rho,\iota)$ is defined by using the spin-foam trace (\ref{sftr}), namely
\be {\rm A}(\kappa,\rho,\iota) \ =\ \left(\prod_{{\rm intrinsic\ faces}\ f} {\rm
dim}(\rho_f)\right)\,{\rm Tr}(\kappa,\rho,\iota) \ee
\item To sum with respect to the spin-network histories with the amplitude as a
weight, one fixes an orthonormal basis
in each space ${\rm Inv}_{\rm
Simp}(\rho_1\otimes...\rho_k\otimes\rho^*_{k+1}\otimes...\otimes\rho^*_{N})$ of
the simple intertwiners. In the (suitably defined) sums the intertwiners run
through  the fixed basis, for each choice of the  representations {at each}
vertex.  
   
\end{itemize}

\subsection{Simple intertwiners}\label{Subsec:simpint}
There are three main proposals for the simple intertwiners: 
\begin{enumerate}

\item that of Barrett-Crane (BC) corresponding to the Palatini action, 

\item that of Engle-Pereira-Rovelli-Livine (EPRL) corresponding  the Holst action
with the value of the Barbero-Immirzi parameter $\gamma\not= \pm1$, 

\item that of Freidel-Krasnov (FK) also  corresponding to the Holst action
with the value of the Barbero-Immirzi parameter $\gamma\not= \pm1$,

\end{enumerate}

The BC intertwiners were derived by passing to the quantum level  some symmetry between the self-dual and anti-self dual 2-forms. A few years ago it was
noticed \cite{BC_error}, that the BC states are insufficient to define physical semi-classical states
of quantum gravity. That observation produced a new activity and led to the  EPRL theory. The EPRL approach uses the Holst action. The simplicity constraint is systematically formulated and quantized. Its quantum solutions are derived.
It is often stated that the failure of the BC approach consists in  the too strong
formulation of the quantum constraints. This is in fact not true.
One can just rewrite the EPRL paper \cite{EPRL} replacing the Holst action by that of Palatini and repeating each argument and every calculation. The result will be the BC intertwiners.  Hence, the real difference is in replacing the Palatini action by that of Holst. The FK model also uses the Holst action. The quantum simplicity constraint is encoded in a  construction of suitable  kinematical semiclassical states. Remarkably, the result coincides with the EPRL model as long as $0<\gamma<1$, whereas the result is different for $\gamma>1$. In particular $\gamma\rightarrow \infty$ is no longer the BC model \cite{engle}.

Below we start with the definition of the n-valent BC intertwiners which basically is equivalent to that of \cite{BC_nvalent,BC_R,BC_O}.  Next we introduce the generalized EPRL intertwiners.

We do not study the FK intertwiners in the $\gamma>1$ in this paper.
         
\subsubsection{The BC intertwiners}\label{Subsubsec:BC} 
The original BC formulation of the simplicity constraint relies on the fact
that, given a 2-surface $S\subset\Sigma$, the metric area  can be expressed in
two equivalent ways by the self-dual / anti-self dual part 
$$ E_{IJ}^\pm\ =\ \frac{1}{2}(E^{IJ}\pm\epsilon_{IJKL}E^{KL}) $$
of the frame field,
\be  {\rm Ar}_S(E^+)\ =\ {\rm Ar}_S(E^-). \ee
The corresponding quantum area operators are defined in the Hilbert space
L$^2({\cal A}^+(\Sigma),{\rm d}\mu_0)\otimes$ L$^2({\cal A}^-(\Sigma),{\rm d}\mu_0)$,
and take the form
\be {\rm Ar}_S(\hat{E}^+)\ =\ \widehat{\rm Ar}_S\otimes 1,\ \ \ \ \ \ \ \ \ \ 
{\rm Ar}_S(\hat{E}^-)\ =\ 1\otimes\widehat{\rm Ar}_S, \ee
where $\widehat{\rm Ar}_S$ is the usual LQG area operator.  The BC quantum  simplicity constraint is 
\be (\widehat{\rm Ar}_S\otimes 1)\,\Psi\ =\ (1\otimes\widehat{\rm Ar}_S)\,\Psi \ee 
for every 2-surface $S$. This set of equations labelled by various 2-surfaces $S$, passes, via the spin-network cylindrical functions,  to  sets of equations defined in the spaces of the intertwiners between  the tensor products of the representations of SU(2)$\times$SU(2). It is sufficient to consider the invariants of the tensor products.                   
      
A representation $\rho_{j^+j^-}$ of SU$(2)\times$SU$(2)$ is a pair of SU$(2)$ representations
$(\rho_{j^+},\rho_{j^-})$, $j^+,j^-=0,\frac{1}{2},...,\frac{n}{2},...$ acting  in the tensor product ${\cal H}_{j^+}\otimes{\cal H}_{j^-}$
\be \rho_{j^+j^-}(g^+,g^-)\ =\ \rho_{j^+}(g^+)\otimes\rho_{j^-}(g^-).  \ee

Given the tensor product $\rho_{j^+_1j^-_1}\otimes\,...\,\otimes\rho_{j^+_nj^-_n}$
the space of the invariants ( {intertwiners of the tensor product
of
representations into the trivial representation}) can be viewed as
\begin{equation} \begin{split} {\rm Inv}\left( \rho_{j^+_1j^-_2}\otimes\,...\,\otimes\rho_{j^+_nj^-_n} \right)\ =\ {\rm Inv}\left( \rho_{j^+_1}\otimes\,...\,\otimes\rho_{j^+_n} \right)\,\otimes\,
{\rm Inv}\left( \rho_{j^-_1}\otimes\,...\,\otimes\rho_{j^-_n} \right)\ \\ \subset\ 
{\cal H}_{j^+_1}\otimes\,...\,\otimes{\cal H}_{j^+_n}\,\otimes\,
{\cal H}_{j^-_1}\otimes\,...\,\otimes{\cal H}_{j^-_n}. \end{split} \end{equation}
In each space ${\rm Inv}\left( \rho_{j_1}\otimes\,...\,\otimes\rho_{j_n}\right)
$ of the SU(2) intertwiners,  the LQG area operators {defined by all
surfaces passing through the vertex} induce a family of operators.  {There
are only finitely many distinct among them.} Slightly abusing the notation let
us  denote the induced area operators by $\widehat{\rm
Ar}_{(1)}^{\rho},...,\widehat{\rm Ar}_{(m)}^{\rho}$, where 
$\rho=(\rho_{j_1},\,...\, ,\rho_{j_n})$. 
 The BC  condition passes now to
\be (\widehat{\rm Ar}_{(I)}^{(\rho^+)}\otimes 1)\,\iota\ =\ (1\otimes\widehat{\rm Ar}_{(I)}^{(\rho^-)})\,\iota \ee 
in the tensor product ${\rm Inv}\left( \rho_{j^+_1}\otimes\,...\,\otimes\rho_{j^+_n} \right)\,\otimes\,{\rm Inv}\left( \rho_{j^-_1}\otimes\,...\,\otimes\rho_{j^-_n} \right)$. Solutions to this equation can be called the BC intertwiners. 

In the index notation the BC condition reads
 \be  \widehat{\rm
Ar}^{(\rho^+)A_1^+...A_n^+}_{(I)}{}_{C_1^+...C_n^+}\,\iota^{
C^+_1...C^+_nA^-_1...A^-_n}\ =\ 
  \widehat{\rm Ar}^{(\rho^-)A^-_1...A^-_n}_{(I)}{}_{C^-_1...C^-_n}\,\iota^{A^+_1...A^+_nC^-_1...C^-_n}.
   \label{BCcond}\ee
   
Among the induced area operators there are $\widehat{\rm Ar}_{(I)}^{(\rho)}$,
$I=1,...,n$ {(we have ordered them to be the first ones)} such that
\be \widehat{\rm Ar}_{(I)}^{(\rho^+)}\otimes 1\ = \
{\sqrt{j^+_I(j^+_I+1)}}(1\otimes 1),\ \ \ \ \ 1\otimes\widehat{\rm
Ar}_{(I)}^{(\rho^-)}\ =\ {\sqrt{j^-_I(j^-_I+1)}}(1\otimes 1). \ee       
It follows that a necessary condition for the existence of a BC intertwiner in
a given intertwiner space is
\be j^+_1\ =\ j^-_1,\
 \ \ \ \ \ \ \ \ \ \ \ \ j^+_n\ =\ j^-_n\,.\ee
Hence,  each SU(2)$\times$SU(2) intertwiner space which contains a BC intertwiner has the form  
 $${\rm Inv}\left( \rho_{j_1}\otimes\,...\,\otimes\rho_{j_n} \right)\,\otimes\,{\rm Inv}\left( \rho_{j_1}\otimes\,...\,\otimes\rho_{j_n} \right).$$ 
Now, the BC condition is (we drop the  decoration $\rho$ of the operators, because now
it is the same on the left and on the right-hand side)
\be (\widehat{\rm Ar}_{(I)}\otimes 1)\,\iota\ =\ (1\otimes\widehat{\rm Ar}_{(I)})\,\iota \label{BCcond2} \ee
or in the index notation
\be
\widehat{\rm Ar}^{A_1^+...A_n^+}_{(I)}{}_{C_1^+...C_n^+}\,\iota^{C^+_1...C^+_nA^-_1...A^-_n}\ =\ 
 \widehat{\rm Ar}^{A^-_1...A^-_n}_{(I)}{}_{C^-_1...C^-_n}\,\iota^{A^+_1...A^+_nC^-_1...C^-_n}.\ee 
 {This} induced BC condition means that $\iota$ is
a bilinear 2-form, the operators  $\widehat{\rm Ar}_{(I)}$
are symmetric with respect to.  

On the other hand, there is a distinguished bilinear form, which is an element
of the intertwiner space at the 
 $${\rm Inv}\left( \rho_{j_1}\otimes\,...\,\otimes\rho_{j_n} \right)\,\otimes\,{\rm Inv}\left( \rho_{j_1}\otimes\,...\,\otimes\rho_{j_n} \right)$$
defined by the unique invariants 
\be \epsilon_{j}\ \in\ {\rm Inv}(\rho_j\otimes\rho_j).    \ee
The normalization and reality condition reduce the rescaling ambiguity to
$\pm 1$. The product gives a distinguished non-invariant element  
\be \epsilon\ :=\ \epsilon_{j_1}\otimes\,...\,\otimes\epsilon_{j_n}\ \in\ {\cal H}_{j_1}\otimes\,...\,\otimes{\cal H}_{j_n}\,\otimes\, 
{\cal H}_{j_1}\otimes\,...\,\otimes{\cal H}_{j_n}\,,\ee 
and the projection onto the intertwiner space,
\be P\ : {\cal H}_{j_1}\otimes\,...\,\otimes{\cal H}_{j_n}\ \rightarrow\
 {\rm Inv}\left( \rho_{j_1}\otimes\,...\,\otimes\rho_{j_n} \right) \label{P}\ee
 gives the distinguished intertwiner, 
\be \iota_{\rm BC}\ :=\ (P\otimes P)\epsilon\, .\ee 

Now, let us see that $\iota_{\rm BC}$ satisfies the induced  BC constraint (\ref{BCcond2}). 
Every operator $L$ in ${\rm Inv}\left( \rho_{j_1}\otimes\,...\,\otimes\rho_{j_n} \right)$ that  commutes with the projection $P$, and is symmetric with respect to  the bilinear form $\epsilon$ satisfies
\be (L\otimes 1)(P\otimes P)\epsilon\ =\ (P\otimes P)(L\otimes 1)\epsilon\ =\ (P\otimes P)(1\otimes L)\epsilon\ =\ (1\otimes L)(P\otimes P)\epsilon.\ee  
In particular the induced area operators have those properties, hence indeed
$\iota_{\rm BC}$ does satisfy the BC condition (\ref{BCcond2}).  

{The only thing one should check is that the intertwiner $\iota_{\rm BC}$ is not
zero.}
 The non-vanishing of $\iota_{\rm BC}$ whenever the SU(2) intertwiner space
${\rm Inv}\left( \rho_{j_1}\otimes\,...\,\otimes\rho_{j_n} \right)$ is not empty
itself, follows from the nondegeneracy of the invariant antisymmetric  bilinear
form   $\epsilon_{\frac{1}{2}}$ which implies the nondegeneracy of each
$\epsilon_{j}$, as well as the nondegeneracy of the product  $\epsilon$. The
non-vanishing of the
 projection onto the intertwiner subspace is the result of  the identity
 \be (1\otimes P) \epsilon\ =\ (P\otimes P)\epsilon\ =\ (P\otimes 1)\epsilon\, .\ee
In words, the identity means that if we apply the projection once, that is 
consider say   $(1\otimes P) \epsilon$, then the result is already an invariant
with respect to the action of  $\rho_{j_1}\otimes\,...\,\otimes\rho_{j_n} $
in the first factor of the tensor product.
Perhaps it is easier to see the non-vanishing in the index notation. According
to the identity,
\be \iota_{\rm BC}{}^{A_1...A_nB_1...B_n}\ =\ P{}^{A_1...A_n}{}_{C_1...C_n}\epsilon_{j_1}^{C_1B_1}...\epsilon_{j_n}^{C_nB_n}. \ee 
But the raising of each index $B_i$ with  $\epsilon_{j_i}$  can not kill any nonzero tensor, in particular the projection $P$.
(One may briefly say that $\iota_{\rm BC}$ is the identity operator 
${\rm Inv}\left( \rho_{j_1}\otimes\,...\,\otimes\rho_{j_n} \right)\ \rightarrow\ {\rm Inv}\left( \rho_{j_1}\otimes\,...\,\otimes\rho_{j_n} \right)$
with the lower indices raised by the invariant bilinear form $\epsilon$.) 
Remarkably, if we rewrite the EPRL paper \cite{EPRL} for the Palatini action (rather
the that of Holst considered by EPRL), we will derive for every intertwiner space
${\rm Inv}\left( \rho_{j_1}\otimes\,...\,\otimes\rho_{j_n} \right)\,\otimes\,{\rm Inv}\left( \rho_{j_1}\otimes\,...\,\otimes\rho_{j_n} \right)$ a unique solution.
This solution will be exactly our very $\iota_{\rm BC}$ (see also below).  

\subsubsection{The EPRL intertwiners}\label{Subsubsec:EPRL} The simplicity constraint (\ref{simpl}) defined by the Holst action depends on a real parameter $\gamma\not=0$, known as the Barbero-Immirzi parameter \cite{barbero-immirzi}.  As in the BC case,   a quantization of the Holst simplicity constraint passes to the intertwiner spaces  
\be {\rm Inv}\left( \rho_{j^+_1j^-_2}\otimes\,...\,\otimes\rho_{j^+_nj^-_n} \right)\ =\ {\rm Inv}\left( \rho_{j^+_1}\otimes\,...\,\otimes\rho_{j^+_n} \right)\,\otimes\,
{\rm Inv}\left( \rho_{j^-_1}\otimes\,...\,\otimes\rho_{j^-_n} \right).\ee 
According to the EPRL quantum simplicity constraint  a necessary condition for the existence of a solution in a given intertwiner space  is  
\be j^+_I\ =\ \frac{|\gamma+1|}{|\gamma-1|}j^-_I,\ \ \ \ \ \ \ I=1,...,n\, .
\label{j+j-}\ee       
An ingredient of  the EPRL intertwiner is an  SU(2) invariant  
\be {\omicron}\ \in {\rm Inv}\left( \rho_{k_1}\otimes\,...\,\otimes\rho_{k_n} \right), \label{omicron}\ee  
where, to every pair $j^+_I,j^-_I$ such that (\ref{j+j-}) the EPRL quantum constraint condition boils down to selecting a representation
$\rho_{k_I}$ of SU(2), of the spin $k_I$ adjusted as follows
\be k_I\ :=\ \begin{cases}  j^++j^-, \ \ {\rm if\ } -1<\gamma<1  \\
                           |j^+-j^-|,\ \ \ {\rm if\ } \gamma<-1\ \ {\rm or\ } 1<\gamma \,. \end{cases}\label{k}\ee
Given $\omicron$, an EPRL intertwiner $\iota_{\rm EPRL}(\omicron)$ is defined using              the invariants
\be c_I\ \in {\rm Inv}\left(\rho_{j^+_I}\otimes\rho_{j^-_I}\otimes\rho^*_{k_I}\right), \ \ \ \ \ \ \ I=1,...,n \label{c}\ee
unique for every $I=1,...,n$, modulo a factor (the normalization and reality can restrict the ambiguity
to $\pm 1$). First, construct a non-invariant element of ${\cal H}_{j^+_1}\otimes\, ...\,\otimes{\cal H}_{j^+_n}\,\otimes\,{\cal H}_{j^-_1}\otimes\, ...\,\otimes{\cal H}_{j^-_n}$, namely
\be  c_1\otimes\,...\,\otimes c_n \lrcorner \omicron\, \ee
that is, in the index notation
\be c_1{}^{A_1^+A_1^-}{}_{B_1}\,...\,c_n{}^{A_n^+A_n^-}{}_{B_n}\omicron^{B_1...B_n}. \ee        
Next, use the projections   
$$ P^\pm\ :\ {\cal H}_{j^\pm_1}\otimes\, ...\,\otimes{\cal H}_{j^\pm_n}\ \rightarrow\ 
{\rm Inv}\left( {\rho}_{j^\pm_1}\otimes\, ...\,\otimes{\rho}_{j^\pm_n}\right),$$ 
that is define
\be\iota_{\rm EPRL}(\omicron)\ :=\ (P^+\otimes P^-)c_1\otimes\,...\,\otimes c_n \lrcorner \omicron.\ee

This formula can be written in a simpler way by skipping one of the projections,
namely 
\be\iota_{\rm EPRL}(\omicron)\ =\ (P^+\otimes 1)\,c_1\otimes\,...\,\otimes c_n \lrcorner \omicron\ =\ (1\otimes P^-)\,c_1\otimes\,...\,\otimes c_n \lrcorner \omicron\,,\ee 
and in the index notation
\be \iota_{\rm EPRL}(\omicron)^{A^+_1...A^+_nA^-_1...A^-_n}\ =\ P^+{}^{A^+_1...A^+_n}_{D^+_1...D^+_n}\,
c_1{}^{D_1^+A_1^-}{}_{B_1}\,...\,c_n{}^{D_n^+A_n^-}{}_{B_n}\,
\omicron^{B_1...B_n}\label{EPRLindex} \ee
As in the BC case, one can briefly describe that definition by saying that 
each lower index of the projection $P^\pm$ is raised by the bilinear form
$c_1{}^{D_1^+A_1^-}{}_{B_1}\,...\,c_n{}^{D_n^+A_n^-}{}_{B_n}\,\omicron^{B_1...B_n}$.

Notice, that in the limit $\gamma\rightarrow \pm \infty$ the conditions 
(\ref{j+j-},\ref{omicron},\ref{k},\ref{c},\ref{EPRLindex}) 
go to (see Section \ref{Subsec:simpint})
\begin{gather} 
j^+_I\ =\ j^-_I,\ \ k_I=0,\ \ c_I\ =\ \epsilon_{j_I},\ \ \omicron\ =\ 1 \\ \iota_{\rm EPRL}(1)^{A^+_1...A^+_nA^-_1...A^-_n}\ :=\ P^+{}^{A^+_1...A^+_n}_{D^+_1...D^+_n}\,
\epsilon_{j_1}{}^{D_1^+A_1^-}\,...\,\epsilon_{j_n}{}^{D_n^+A_n^-}\ =\
{\iota_{\rm BC}}^{A^+_1...A^+_nA^-_1...A^-_n}.
\end{gather} 
That is, the EPRL intertwiner becomes the BC intertwiner $\iota_{\rm BC}$.  
This is consistent with the fact that the classical Holst action  becomes the
Palatini action  in the limit $\gamma\rightarrow \pm\infty$.
        
In practical calculations, the projector (\ref{P}) can be written in terms of any basis 
\be \iota_1,...,\iota_N\ \in\ {\rm Inv}\left(\rho_{j_1}\otimes...\otimes\rho_{j_n} \right),
\ee
namely 
\be P^{A_1...A_n}_{B_1...B_n}\ =\ \sum_{i=1}^N \iota_i^{A_1...A_n}\iota^{i*}_{B_1...B_n} \ee 
where $\iota^{1*}, ..., \iota^{N*}$ is the dual basis. 
In particular, one can use an orthonormal basis, and then
$$\iota^{i*}\ =\ \iota_i^\dagger.$$

The $3j$, $6j$, $10j$ and $15j$  symbols emerging in the context of the
BC and EPRL intertwiners are related to specific choice of basis 
in the valency 4 of the spin-networks and  to the 4-tetrahedral spin-foams.
The advantage of our notation is the independence of the choice of basis.

\subsubsection{The injectivity of $\omicron\mapsto \iota_{\rm
EPRL}(\omicron)$
in the case $\gamma\ge 1$}
As  one could see above, the EPRL spin-network SU(2)$\times$SU(2)
intertwiners $\iota_{\rm EPRL}(\omicron)$ are labelled by the SU(2)
intertwiners $\omicron$.
However, we have not said anything about the kernel of the map
$$\omicron\ \mapsto\ \iota_{\rm EPRL}(\omicron).$$
We will show now that this map is injective in the case
$$k_I\ =\ j^+_I\ -\ j^-_I\, ,$$ provided $j^-_1+\ldots+j^-_n\in \mathbb{N}$.

Suppose we have found some $\omicron_0$, such that
\be \iota_{\rm EPRL}(\omicron_0)\ =\ 0,\ee
 that is
\be P^+{}^{A^+_1...A^+_n}_{D^+_1...D^+_n}\,
c_1{}^{D_1^+A_1^-}{}_{B_1}\,...\,c_n{}^{D_n^+A_n^-}{}_{B_n}\,
\omicron_0{}^{B_1...B_n}\ =\ 0.\label{PomicronW}\ee
Then, any contraction of the left-hand side also vanishes, in particular
\be P^+{}^{A^+_1...A^+_n}_{D^+_1...D^+_n}\,
c_1{}^{D_1^+}{}_{A_1^-B_1}\,...\,c_n{}^{D_n^+}{}_{A_n^-B_n}\,
\omicron_0{}^{B_1...B_n}W^{A_1^-...A_n^-}\ =\ 0, \ee
where we lowered  the indices $A^-_I$  with the $\epsilon_{j^-_I}$
bilinear forms
and took any $0\not= W\in {\rm
Inv}\left(\rho_{j^-_1}\otimes\,...\,\otimes\rho_{j^-_n}\right)$. Such $W$ exists iff $j^-_1+\ldots+j^-_n\in \mathbb{N}$.
We will show now, that the last equality can not be true  unless
$\omicron_0 = 0$ itself.

First, let us realize some  property of the spin composition map
defined given two half integers $k$ and $j$,
\be c_{(k,j)}\ :\ {\cal H}_{k}\otimes{\cal H}_{j}\ \rightarrow {\cal
H}_{k+j}\label{sc} \ee
intertwining the corresponding SU(2) representations. In our notation
the map  is given  by the intertwiners {$c^{A}_{\phantom{A}BC}$}, namely
{\be {\cal
H}_k\otimes{\cal H}_{j}\ni K^{BC}\ \mapsto  K^{BC} c^A_{\phantom{A}BC}\in {\cal
H}_{k+j}.\ee}
The key property is that the map can not annihilate a nonzero simple
tensor { $K^{BC}=\omicron^BW^C$}
\be \omicron^BW^C c^A_{\phantom{A}BC}\ =\ 0\ \ \ \ \Leftrightarrow \
\omicron^BW^C \ =\ 0\label{basic}\ee
To see that this is really true, view each of the Hilbert spaces ${\cal
H}_{l}$, $l\in \frac{1}{2}\mathbb{N}$ as the symmetric part of the tensor
product
${\cal H}_{\frac{1}{2}}\otimes...\otimes{\cal H}_{\frac{1}{2}}$, introduce
an orthonormal  basis $|0 \rangle,|1 \rangle\in {\cal H}_{\frac{1}{2}}$
and
use it to define a basis $v^{(l)}_0,...,v^{(l)}_{2l}\ \in {\cal H}_{l}$,
\be
\begin{array}{rcl}
v^{(l)}_0\ &:= &\ |0 \rangle\otimes\,...\,\otimes|0 \rangle,\nonumber\\
v^{(l)}_{1}\  &:= &\  {\rm Sym}|0 \rangle\otimes\,...\,\otimes|0 \rangle\otimes|1\rangle,\\
&...& \\
v^{(l)}_{2l}\ &:= &\ |1 \rangle\otimes\,...\,\otimes|1 \rangle,
\end{array}
\ee
where 
$$ {\rm Sym}: {\cal H}_{\frac{1}{2}}\otimes\,...\,{\cal H}_{\frac{1}{2}}\ \rightarrow\  {\cal H}_{\frac{1}{2}}\otimes\,...\,{\cal H}_{\frac{1}{2}}$$
denotes the projection on the subspace of symmetric tensors.  The
spin composition map in this basis reads
\be c_{(k,j)}\ :\ v^{(k)}_m\otimes v^{(j)}_{m'}\ \mapsto\
v^{(k+j)}_{m+m'}. \ee
Hence, the product of two general elements is mapped in the following way
\be \sum_{m=m_1}^{2k}\omicron'^{m}v^{(k)}_m\,\otimes\,
\sum_{m'=m'_1}^{2j}W'^{m'}v^{(j)}_{m'}\ \mapsto\
\omicron'^{m_1}W'^{m'_1}v^{(k+j)}_{m_1+m'_1}\ +\
\sum_{M>m_1+m'_1}\alpha^Mv^{(k+j)}_M\ \not=\ 0,\label{spincomp}  \ee
where $\omicron^{m_1}$ and $W^{m'_1}$ are the first non-vanishing
components.
The result can not be zero, because the first term on the right-hand side
is nonzero,
and is linearly independent of the remaining terms.

Secondly, we generalize the statement (\ref{basic})  to the following one:
\be c_1{}^{D_1^+}{}_{A_1^-B_1}\,...\,c_n{}^{D_n^+}{}_{A_n^-B_n}\,
\omicron{}^{B_1...B_n}W^{A_1^-...A_n^-}\ =\ 0 \ \ \Leftrightarrow\
\ \omicron{}^{B_1...B_n}W^{A_1^-...A_n^-}\ =\ 0,\label{gen}\ee
for arbitrary $\omicron\in{\cal H}_{k_1}\otimes...\otimes{\cal H}_{k_n}$
and $W\in {\cal H}_{j^-_1}\otimes...\otimes{\cal H}_{j^-_n}$.
In the proof we will use a calculation similar to that of
(\ref{spincomp}), with the difference that
 now the coefficients in (\ref{spincomp}) take values in the $n-1$ valent
tensor products. Specifically,
the left-hand side of the first equality in (\ref{gen}) is the result
of the map
\be c_{(k_1,j^-_1)}\otimes...\otimes c_{(k_n,j^-_n)}\ :\ {\cal
H}_{k_1}\otimes\,...\,\otimes{\cal H}_{k_n}\,\otimes\,{\cal
H}_{j^-_1}\otimes\,...\,\otimes{\cal H}_{j^-_n}\ \rightarrow\
{\cal H}_{k_1+j^-_1}\otimes\,...\,\otimes {\cal H}_{k_n+j^-_n},\label{msc}\ee
 defined by (\ref{sc}), and applied to  given $\omicron'$ and $W'$. By
analogy
 to (\ref{spincomp}), we can write
 \be \omicron\ =\ \sum_{m=m_1}^{2k_1}v^{(k_1)}_m\otimes
 \omicron^{m},\ \ \ \ W\ =\ \sum_{m^-=m^-_1}^{2j^-_1}v^{(j^-_1)}_{m^-}
\otimes W^{m^-}\ee
where $\omicron^{m_1}$ is the first nonvanishing ${\cal
H}_{k_2}\otimes...\otimes{\cal H}_{k_n}$ valued component
of $\omicron$ and $ W^{m^-_1}$ is the first nonvanishing ${\cal
H}_{j^-_2}\otimes...\otimes{\cal H}_{j^-_n}$ valued component
of $W$. Now we apply the map (\ref{msc}),
 \begin{equation} \label{mspincomp} \begin{split} c_{(k_1,j^-_1)}\otimes...\otimes c_{(k_n,j^-_n)}\,
&\left(\sum_{m=m_1}^{2k_1}v^{(k_1)}_m\otimes
 \omicron^{m}\,\otimes\,\sum_{m^-=m^-_1}^{2j^-_1}v^{(j^-_1)}_{m^-}
\otimes W^{m^-}\right)\\
\ &=\ v^{(k_1+j^-_1)}_{m_1+m^-_1}\otimes
\left(c_{(k_2,j^-_2)}\otimes...\otimes
c_{(k_n,j^-_n)}\left(\omicron^{m_1}\otimes W^{m^-_1}\right)\right)\ +\\
&+\ \sum_{M>m_1+m^-_1}v^{(k_1+j^-_1)}_M\otimes \alpha^M.
\end{split}
\end{equation}

The first term on the right-hand side is linearly independent of the
others, hence if it were
nonzero so would be the right-hand side. But it is nonzero provided
(\ref{gen}) holds for $n$ replaced by $n-1$. Since (\ref{gen}) is true
for $n=1$, the mathematical induction does the rest of the job.

Finally, we notice that if
$$\omicron=\omicron_0\in {\rm
Inv}({\rho}_{k_1}\otimes\,..,\otimes{\rho}_{k_n})$$ and
$$W \in {\rm Inv}({\rho}_{j^-_1}\otimes\,..,\otimes{\rho}_{j^-_n}),$$
then the tensor
\be c_1{}^{D_1^+}{}_{A_1^-B_1}\,...\,c_n{}^{D_n^+}{}_{A_n^-B_n}\,
\omicron_0{}^{B_1...B_n}W^{A_1^-...A_n^-} \ee
defines an element of   ${\rm
Inv}({\rho}_{j^+_1}\otimes\,..,\otimes{\rho}_{j^+_n})$,
that is,
\be c_{(k_1,j^-_1)}\otimes...\otimes c_{(k_n,j^-_n)}(\omicron_0\otimes W)\
\in
 {\rm Inv}({\rho}_{j^+_1}\otimes\,..,\otimes{\rho}_{j^+_n}).\ee

Hence the projection $P^+$ in (\ref{PomicronW}) acts as the identity,
\be P^+c_{(k_1,j^-_1)}\otimes...\otimes c_{(k_n,j^-_n)}(\omicron_0\otimes
W)\ =\
c_{(k_1,j^-_1)}\otimes...\otimes c_{(k_n,j^-_n)}(\omicron_0\otimes W),\ee
 hence
\be
P^+{}^{A^+_1...A^+_n}_{D^+_1...D^+_n}\,c_1{}^{D_1^+}{}_{A_1^-B_1}\,...\,c_n{}^{D_n^+}{}_{A_n^-B_n}\,
\omicron_0{}^{B_1...B_n}W^{A_1^-...A_n^-}\ =\ 0 \ \ \Leftrightarrow\
\ \omicron_0{}^{B_1...B_n}W^{A_1^-...A_n^-}\ =\ 0,\ee
 but $W$ is arbitrary in (\ref{PomicronW}), hence
 $$ \omicron_0\ =\ 0.$$

\section{Discussion}
\subsection{Summary}
The conclusion coming from our work is that we do not have to reformulate
LQG in terms of the piecewise linear category and triangulations to match
it with the EPRL SFM. The generalized spin-foam framework  is
compatible with the original LQG framework and accommodates all the spin-network states and the diffeomorphism covariance.  Our generalization goes in two directions.
The first one is from spin-foams defined on simplicial complexes to
spin-foams defined on the arbitrary linear 2-cell complexes (see also  
\cite{Reisenberger}). The main tools of the SF models needed to describe 4-dimensional spacetime 
are available: spin-foam, boundary spin-network, characterization of  vertex,  vertex
amplitude, the scheme of the  SF models of 4-dim gravity, the EPRL
intertwiners.

The second direction is from abstract spin-foams to embedded spin-foams,
histories
of the embedded spin-network.\footnote{As a matter of fact the simplicial
spin-foams considered in \cite{Baezintro} were also embedded, however they
did not use all the diversity given by the embeddings, since they were
restricted to triangulations.}  For example the notion of knots and
links is again available in that framework.
One may for example consider a spin-foam history which turns an unknotted
embedded circle into a knotted circle.

The most important result is the characterization of a vertex of a
generalized
spin-foam.  The structure of each vertex can be completely encoded in  a
spin-network
induced locally on the boundary of the vertex neighbourhood. The
spin-network is used for the natural generalization of the vertex
amplitude used in the simplicial spin-foam models. The characterization of
the vertices leads to a general construction
of a general spin-foam. The set of all the possible vertices is given by
the set
of the spin-networks and and set of all spin-foams can be obtained by
gluing the vertices and the ``static'' spin-foams.

We have also proved the 1-1 correspondence between the EPRL
SU(2)$\times$SU(2) intertwiners and all the SU(2) intertwiners in the case
of the Immirzi-Barbero parameter $|\gamma| \geq 1$ and non-trivial co-domain of EPRL map.

In the literature, the $3j$, $6j$, $10j$ and $15j$  symbols are used
extensively in the context of the BC and EPRL intertwiners and amplitudes.
From our point of view, they are just related to {a} specific choice of the
basis
in the space of intertwiners
valid in the case of the simplicial spin-foams and spin-networks.
Our approach is basis independent.

\subsection{The limits $\gamma=\pm\infty,\pm 1,0$}
A subject that deserves an individual discussion are the limits of the
EPRL model
as $\gamma\rightarrow 0,\pm 1, \pm\infty$. Interestingly, all those limits
are mathematically well defined, although the model seems to be losing
its
key properties.

The best established limit is
\be \gamma\ \rightarrow\ \pm\infty\,. \ee
That limit exists at each level: the Holst action converges to the
Palatini action,
$j^+=j^-$, $k=0$, the EPRL intertwiner converges to the BC intertwiner,
all the EPRL derivation converges to a finite limit. The only surprise is
the discontinuity in the
number of the degrees of freedom. The BC intertwiner space has $0$ or $1$
dimension per vertex and has been proven to insufficiently accommodate all
the gravitational
degrees of freedom.

The limit that can not be extended to the entire derivation although the
Holst action does converge perfectly well to the self dual action is
\be \gamma\ =\ \pm 1.\ee
However the EPRL intertwiner has a limit in that case:
$$j^\mp=0, k=j^\pm\, ,$$
and moreover
$$ \iota_{\rm EPRL}(\omicron)\ =\ \omicron, $$
where $\omicron$ is an arbitrary SU(2) intertwiner, and the amplitude
turns into the SU(2) BF amplitude. So the limit theory is the SU(2) BF
theory. That is very strange, taking into account that the self dual
action still defines the same Einstein's (Euclidean) gravity.

The limit in which the Holst action is no longer equivalent to the
Palatini action
and (upon the rescaling by $\gamma$) defines {a topological}
theory is
\be \gamma\ =\ 0\,.\ee
Then
\be j^+\ =\ j^-, \ \ \ k\ =\ j^+\ +\ j^- \ee
but, quite surprisingly, the EPRL theory does not resemble {a topological}
theory at all. 
\vskip 0.5cm

\noindent{\bf Acknowledgements} We would like to thank John Barrett, Jonathan Engle
for coming to Warsaw and delivering  lectures on the SFM. JL acknowledges the 
conversations with Laurent Freidel, and exchange of e-mails with John Baez,
Carlo Rovelli, Roberto Pereira and Etera Livine. MK would like to thank Jonathan
Engle from Albert Einstein Institute in Potsdam and Carlo Rovelli, Matteo
Smerlak from Centre de Physique Th\'eorique de Luminy for discussions and
hospitality during his visits at their institutes. The work was partially
supported by the Polish Ministerstwo Nauki i Szkolnictwa Wyzszego grants
182/NQGG/
2008/0,  2007-2010 research project N202 n081 32/1844, the National Science
Foundation (NSF) grant PHY-0456913, by the Foundation for Polish Science grant
Master and a Travel Grant from the QG research networking programme of the European Science Foundation.

\end{document}